\documentclass{PoS}
\usepackage{amsmath,amssymb}

\title{ \vspace{-3em}\begin{flushright}\normalsize \sf MS-TP-17-21\\[3em]\end{flushright}
	Soft gluon resummation for the associated production of a top quark pair with a W boson at the LHC}

\ShortTitle{Soft gluon resummation for the associated production of a top quark pair with a W boson at the LHC}

\author{Anna Kulesza\\
        Institute for Theoretical Physics, WWU M\"unster, D-48149 M\"unster, Germany\\
        E-mail: \email{anna.kulesza@uni-muenster.de}}

\author{Leszek Motyka\\
          Institute of Physics, Jagellonian University, S. \L{}ojasiewicza 11, 30-348 Krak\'ow, Poland\\
        E-mail: \email{leszek.motyka@uj.edu.pl}}

\author{\speaker{Daniel Schwartl\"ander}\\
       Institute for Theoretical Physics, WWU M\"unster, D-48149 M\"unster, Germany\\
       E-mail: \email{d\_schw20@uni-muenster.de}}
        
\author{Tomasz Stebel\\
        Institute of Nuclear Physics PAN, Radzikowskiego 152, 31-342 Krak\'ow, Poland\\
        E-mail: \email{tomasz.stebel@uj.edu.pl}}
        
\author{Vincent Theeuwes\\
          Department of Physics, SUNY Buffalo, 261 Fronczak Hall, Buffalo, NY 14260-1500, USA\\
        E-mail: \email{vtheeuwe@buffalo.edu}}

\abstract{We present our results on soft gluon resummation in the invariant mass threshold applied to the associated production of a top quark pair with a W boson at the LHC in the Mellin space formalism.}

\FullConference{The European Physical Society Conference on High Energy Physics\\
         5-12 July\\
         Venice, Italy}

\begin{document}

\section{Introduction}

The measurements of associated production of a vector boson with a top-antitop quark pair provide an important test for the Standard Model at the LHC \cite{Aaboud:2016xve,Khachatryan:2015sha}. 
In particular these are the key processes to measure the top quark couplings to $W/Z$ bosons. Furthermore they are relevant in searches for new physics due to both being directly sensitive to it and providing an important background.
They also play an important role as a background for the associated Higgs boson production process $pp \rightarrow t \bar t H$.
It is therefore necessary to know the theoretical predictions for $pp \rightarrow t \bar t W/Z$ with high accuracy.

Fixed order cross sections up to next-to-leading order in $\alpha_S$ are already known for some time \cite{Lazopoulos:2008de,Lazopoulos:2007bv}. They were recalculated and matched to parton showers in \cite{Kardos:2011na,Campbell:2012dh,Alwall:2014hca,Garzelli:2011is,Garzelli:2012bn}. Furthermore QCD-EW NLO corrections have been obtained \cite{Frixione:2015zaa}.
While NNLO calculations for this particular type of 2 to 3 processes are currently out of reach, a class of corrections beyond NLO from the emission of soft and/or collinear gluons can be taken into account with the help of resummation methods. 
This was done in the framework of the soft-collinear effective theory (SCET) for $pp \rightarrow t \bar t W$ \cite{Li:2014ula} and with SCET-formulas expressed in Mellin-space for $pp \rightarrow t \bar t W/Z$ \cite{Broggio:2016zgg,Broggio:2017kzi}.

In the following we present results for threshold-resummed cross sections in the invariant mass kinematics, obtained using the Mellin-space approach at NLL accuracy. The calculations are then improved beyond NLL by including non-logarithmic hard contributions of the order ${\cal O}(\alpha_S)$.

\section{Resummation at invariant mass threshold}
Here we treat the soft gluon corrections in the invariant mass kinematics, i.e we consider the limit $\hat \rho = \frac{Q^2}{\hat s} \rightarrow 1$ with $Q^2=(p_t + p_{\bar t} + p_{W/Z})^2$.  
The logarithms resummed in the invariant mass threshold limit have the form
\begin{align}
   \alpha_S^m \left(\frac{\log^n{(1-\hat\rho)}}{1-\hat\rho}\right)_{+} \hspace{25pt} m \le 2n-1
\end{align}
with the plus distribution $\int_0^1\text{d}x (f(x))_{+} = \int_0^1\text{d}x (f(x) - f(x_0))$.
The Mellin moments of the differential cross section $\frac{\text{d} \sigma_{ij \rightarrow t\bar tW/Z}}{\text{d} Q^2}$ are taken with respect to the variable $\rho = \frac{Q^2}{S}$. At the partonic level this leads to
\begin{equation}
 \frac{\text{d} \tilde {\hat \sigma}_{ij \rightarrow t\bar tW/Z}}{\text{d} Q^2} (N,Q^2,m_t,m_{W/Z},\mu_R^2,\mu_F^2)= \int_0^1 \text{d} \hat \rho \hat \rho^{N-1} \frac{\text{d} \hat \sigma_{ij \rightarrow t\bar tW/Z}}{\text{d} Q^2} (\hat \rho,Q^2,m_t,m_{W/Z},\mu_R^2,\mu_F^2)
\end{equation}
for the Mellin moments for the process $ij \rightarrow t\bar tW/Z$ with i,j denoting two massless colored partons. In Mellin space the threshold limit $\hat \rho \rightarrow 1$ corresponds to the limit $N \rightarrow \infty$.
Since the process involves more than 3 colored partons, the resummed cross section involves color matrices. In Mellin space the resummed partonic cross section has the form \cite{Contopanagos:1996nh,Kidonakis:1998nf}
\begin{equation}
 \frac{\text{d} \tilde {\hat \sigma}_{ij \rightarrow t\bar tW/Z}}{\text{d} Q^2} = \text{Tr}[\mathbf{H}_{ij \rightarrow t\bar tW/Z} \mathbf{S}_{ij \rightarrow t\bar tW/Z}] \Delta_i \Delta_j,
\end{equation}
where $\mathbf{H}_{ij \rightarrow t\bar tW/Z}$ and $\mathbf{S}_{ij \rightarrow t\bar tW/Z}$ are color matrices and the trace is taken in color space. We describe the evolution of color in the s-channel color basis, for which the basis vectors are
\begin{equation}
 c_{\mathbf{1}}=\delta_{a_i,a_j}\delta_{a_k,a_l} \quad c_{\mathbf{8}}=T^c_{a_i,a_j}T^c_{a_k,a_l}
\end{equation}
for the $q \bar q$ initial state and 
\begin{equation}
 c_{\mathbf{1}}=\delta_{a_i,a_j}\delta_{a_k,a_l} \quad c_{\mathbf{8S}}=d^{c,a_i,a_j}T^c_{a_k,a_l} \quad c_{\mathbf{8A}}=f^{c,a_i,a_j}T^c_{a_k,a_l}
\end{equation}
for the $gg$ initial state. This choice of color basis leads to a diagonal soft anomalous dimension matrix in the absolute threshold limit $\frac{(2m_t+m_{W/Z})^2}{\hat s} \rightarrow 1$, which is a special case of the invariant mass threshold limit.
$\mathbf{H}_{ij \rightarrow t\bar tW/Z}$ describes the hard scattering contributions projected on the color basis, while $\mathbf{S}_{ij \rightarrow t\bar tW/Z}$ represents the soft wide angle emission. The (soft-)collinear logarithmic contributions form the initial state partons are taken into account by the functions $\Delta_i$ and $\Delta_j$.

At NLL accuracy the evolution of the soft matrix $\mathbf{S}_{ij \rightarrow t\bar tW/Z}$ is given by the one-loop anomalous dimension matrix, see e.g. \cite{Kulesza:2015vda}. Since the soft anomalous dimension matrix is not diagonal in the invariant mass threshold, we use the method proposed in \cite{Kidonakis:1998nf} to diagonalize the soft anomalous dimension matrix in a basis $R$.
Then $\mathbf{S}_{ij \rightarrow t,\bar tW/Z}$ is given by \cite{Kidonakis:1998nf}:
\begin{equation}
 \mathbf{S}_{ij \rightarrow t\bar tW/Z,R} = \mathbf{S}^{(0)}_{ij \rightarrow t\bar tW/Z,R} \exp\left[\int_\mu^{Q/N}\frac{\text{d}q}{q}(\lambda^{*}_{R,I} + \lambda_{R,J})\right]
\end{equation}
where $\lambda_i,\lambda_j,...$ are the eigenvalues of the soft anomalous dimension matrix and $\mathbf{S}^{(0)}_{ij \rightarrow t\bar tW/Z,R}$ is
\begin{equation}
 \left(\mathbf{S}^{(0)}_{ij \rightarrow t\bar tW/Z}\right)_{IJ} = \text{Tr}\left[c^\dagger_I c_J \right]
\end{equation}
transformed into the $R$ basis.
$\mathbf{H}_{ij \rightarrow t\bar tW/Z}$ can be calculated perturbatively 
\begin{equation}
 \mathbf{H}_{ij \rightarrow t\bar tW/Z} = \mathbf{H}^{(0)}_{i,j \rightarrow t,\bar t,W/Z} + \frac{\alpha_S}{\pi}\mathbf{H}^{(1)}_{ij \rightarrow t\bar tW/Z} + ... \, .
\end{equation}
For NLL accuracy it is sufficient to include $\mathbf{H}^{(0)}_{i,j \rightarrow t,\bar t,W/Z}$, which is given by the leading order cross section. To improve the predictions beyond NLL we can also include $\mathbf{H}^{(1)}_{ij \rightarrow t\bar tW/Z}$.
This factor collects non-logarithmic contributions of ${\cal O}(\alpha_S)$ in the large $N$ limit \cite{Beenakker:2011sf,Beenakker:2013mva}. In particular it includes the virtual loop corrections, which are numerically extracted from the PowHel implementation \cite{Garzelli:2011is,Garzelli:2012bn}.
The initial state jet functions $\Delta_i$ and $\Delta_j$ have been known for a long time \cite{Catani:1996yz,Bonciani:1998vc} and depend only on the emitting parton.

\section{Numerical results}
The numerical results were obtained using $m_t=173$\,GeV, $m_W=80.385$\,GeV and MMHT14 PDF sets \cite{pdf} and for the center of mass energy $\sqrt{S}=13$\,TeV. The one-loop hard contributions to $\mathbf{H}^{(1)}_{ij \rightarrow t\bar tW/Z}$ and the NLO cross sections were calculated with the PowHel implementation \cite{Garzelli:2011is,Garzelli:2012bn}. 
We use $\mu = \frac{M}{2}= m_t + \frac{m_W}{2}$ and $\mu = Q$ for the scales $\mu = \mu_R = \mu_F$. Total cross section results were obtained by integrating the resummed differential cross section $\frac{\text{d} \tilde {\sigma}}{\text{d} Q^2}$. These resummed results are then matched to fixed order NLO predictions \cite{Kulesza:2017ukk}.

\begin{figure}
	\centering
	\includegraphics[width=0.75\textwidth]{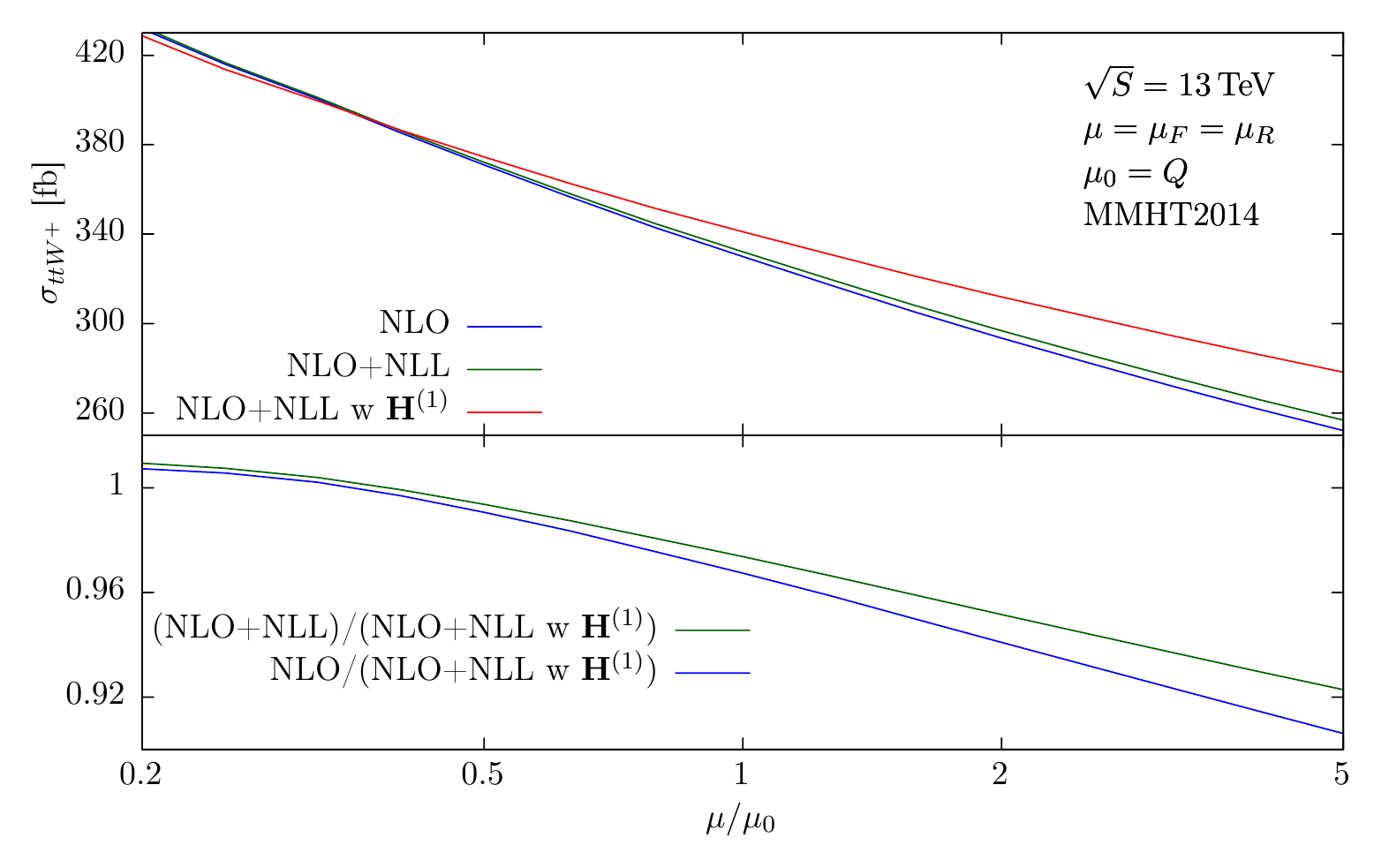}
	\caption{Scale dependence of the total $pp \rightarrow t \bar t W^{+}$ cross section at NLO, NLL matched to NLO and NLL matched to NLO improved with $\mathbf{H}^{(1)}_{ij \rightarrow t\bar tW/Z}$ for the central scale $\mu_0 = Q$.} 
	\label{f:scaledependencettwplusQ}
\end{figure}

\begin{figure}
	\centering
	\includegraphics[width=0.75\textwidth]{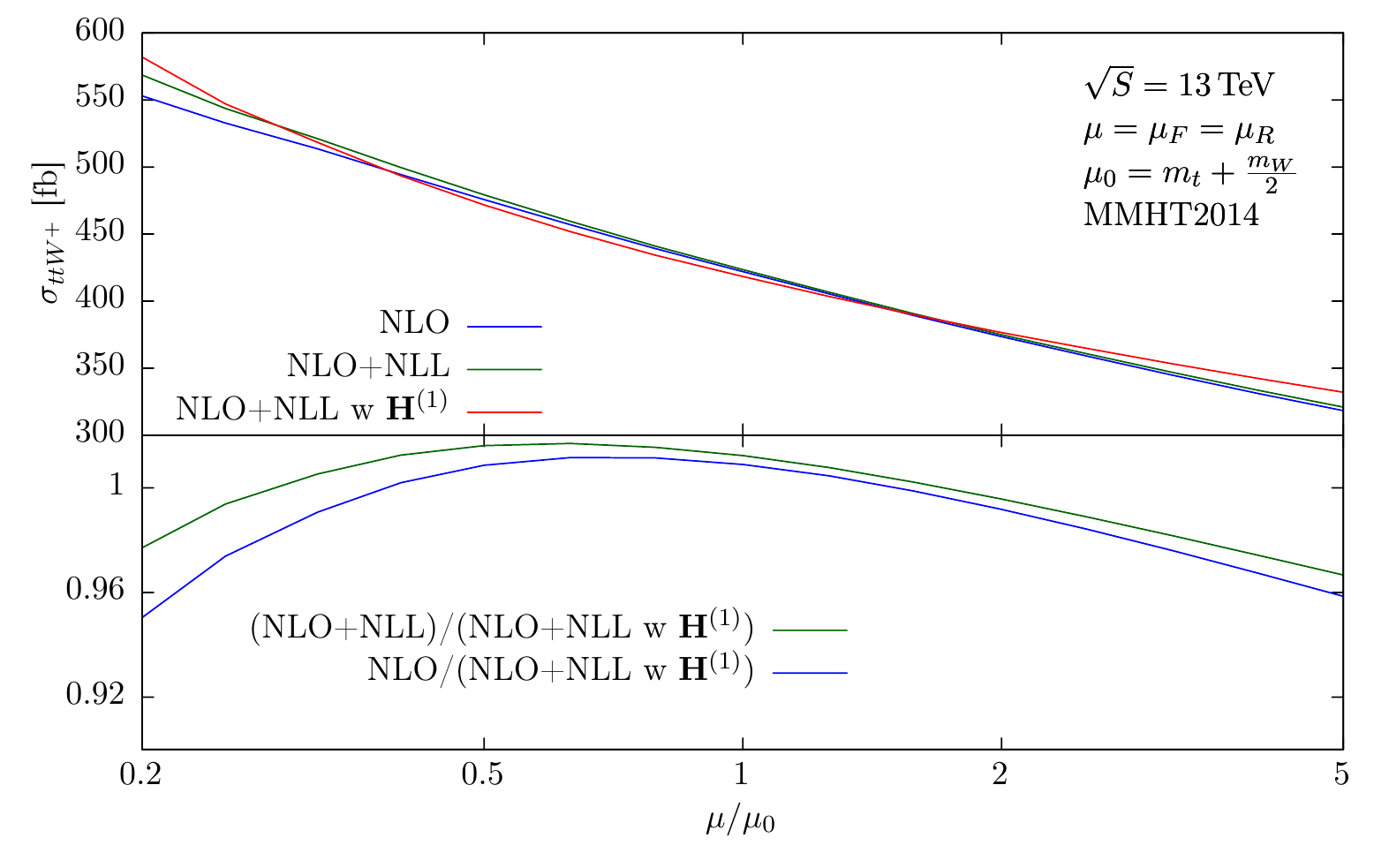}
	\caption{Scale dependence of the total $pp \rightarrow t \bar t W^{+}$ cross section at NLO, NLL matched to NLO and NLL matched to NLO improved with $\mathbf{H}^{(1)}_{ij \rightarrow t\bar tW/Z}$ for the central scale $\mu_0 = \frac{M}{2}$.} 
	\label{f:scaledependencettwplusM}
\end{figure}

\begin{figure}
	\centering
	\includegraphics[width=0.75\textwidth]{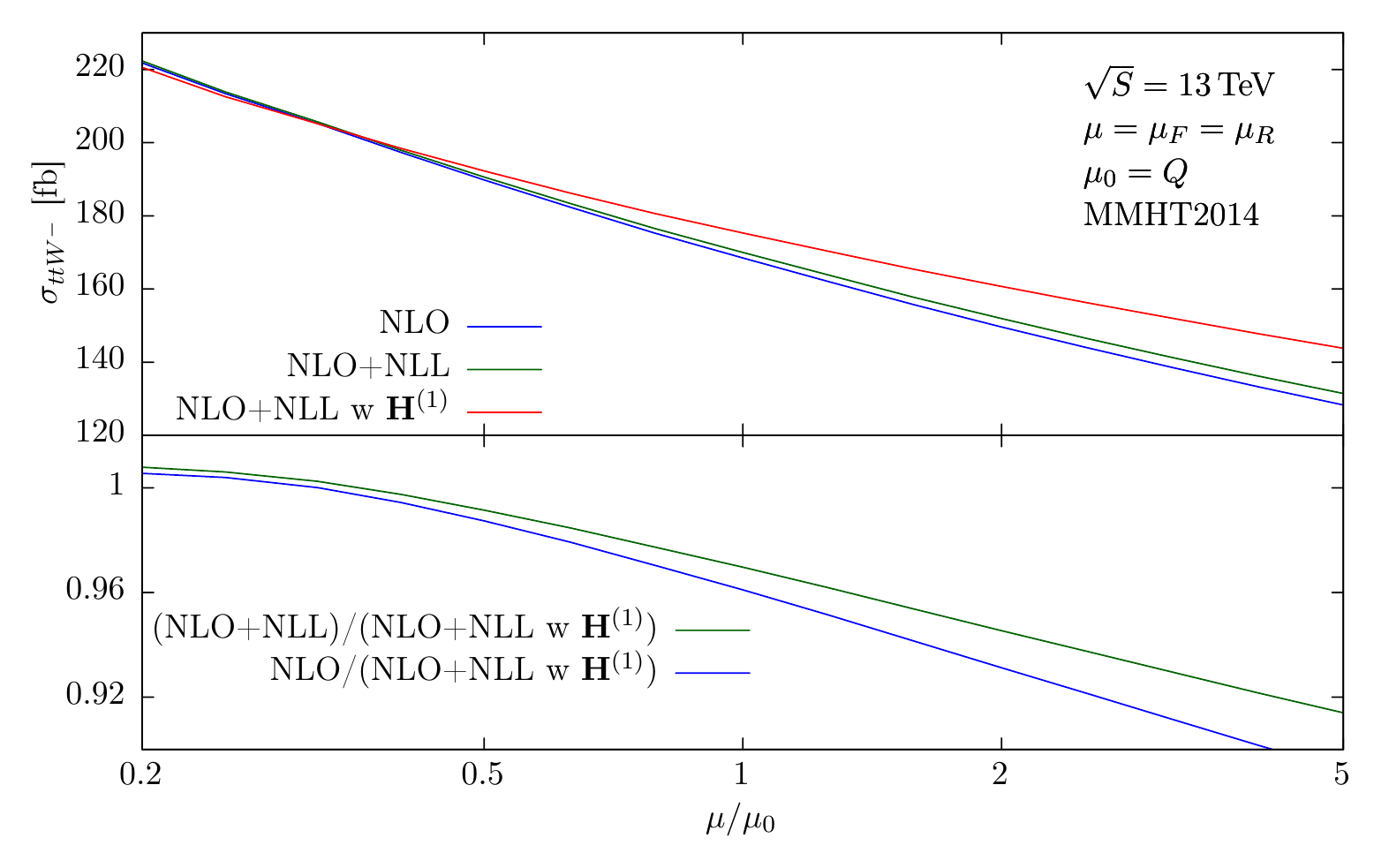}
	\caption{Scale dependence of the total $pp \rightarrow t \bar t W^{-}$ cross section at NLO, NLL matched to NLO and NLL matched to NLO improved with $\mathbf{H}^{(1)}_{ij \rightarrow t\bar tW/Z}$ for the central scale $\mu_0 = Q$.} 
	\label{f:scaledependencettwminusQ}
\end{figure}

\begin{figure}
	\centering
	\includegraphics[width=0.75\textwidth]{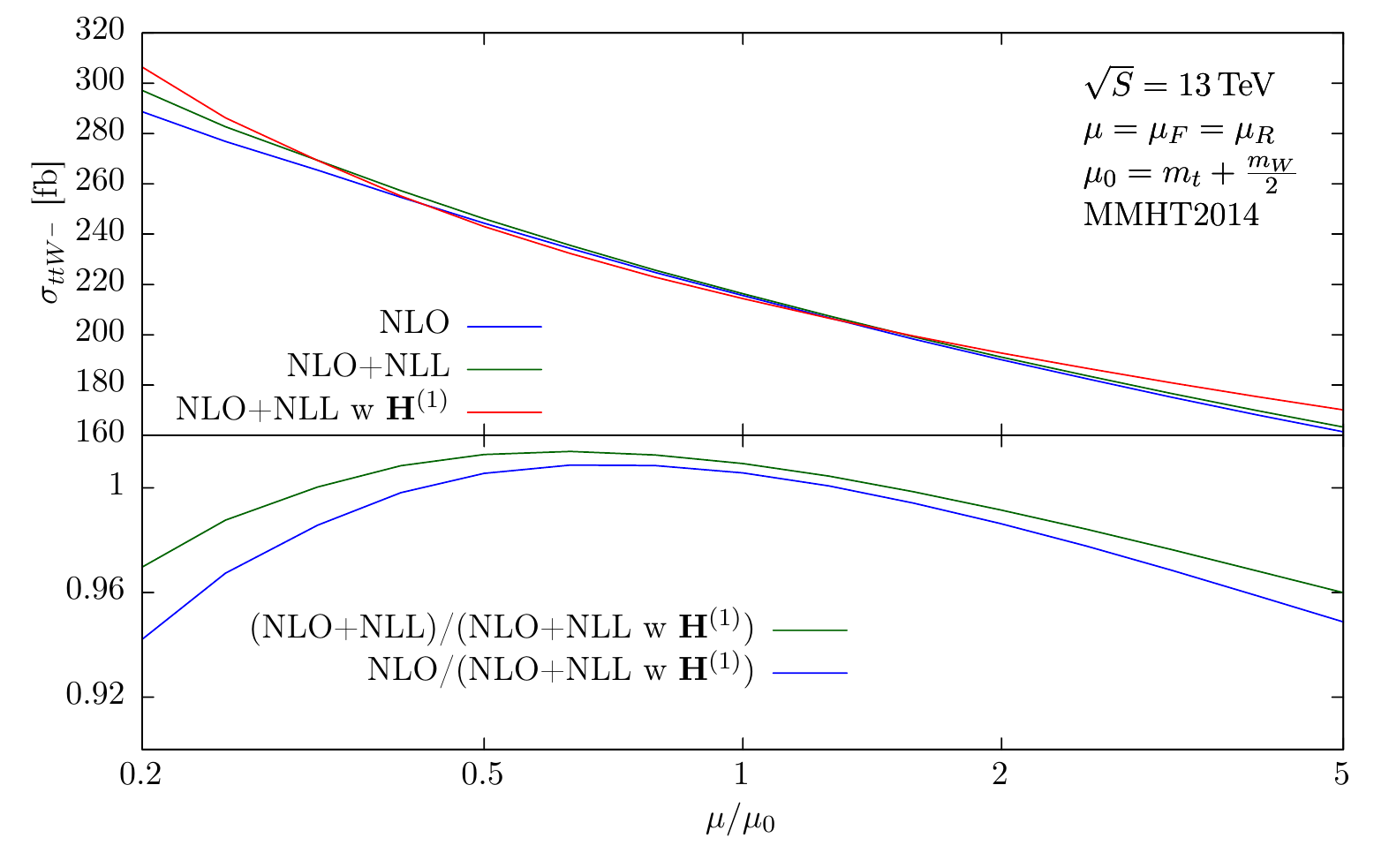}
	\caption{Scale dependence of the total $pp \rightarrow t \bar t W^{-}$ cross section at NLO, NLL matched to NLO and NLL matched to NLO improved with $\mathbf{H}^{(1)}_{ij \rightarrow t\bar tW/Z}$ for the central scale $\mu_0 = \frac{M}{2}$.} 
	\label{f:scaledependencettwminusM}
\end{figure}

In table \ref{tab:xsections} we show the total cross section for $t \bar t W^{+/-}$ production at the two different central scale choices and their scale uncertainty calculated with the seven point method.
Figures \ref{f:scaledependencettwplusQ}, \ref{f:scaledependencettwplusM}, \ref{f:scaledependencettwminusQ} and \ref{f:scaledependencettwminusM} show the scale dependence of the $t \bar t W^{+/-}$ production cross section by varying simultaneously the renormalization and factorization scale $\mu = \mu_R = \mu_F$. In all figures the NLO results are compared with the resummed NLO + NLL and NLO + NLL with $\mathbf{H}^{(1)}_{ij \rightarrow t\bar tW}$ results for two different central scale choices $\mu = Q$ and $\mu = \frac{M}{2}$.
The resummed NLL matched to NLO with $\mathbf{H}^{(1)}_{ij \rightarrow t\bar tW}$ cross section is less sensitive to scale variation as compared to the NLO result.  At large scales the inclusion of $\mathbf{H}^{(1)}_{ij \rightarrow t\bar tW/Z}$ has a bigger impact on the cross section than the logarithmic contributions.
At central scale the cross section is increased by $3.4\%$ ($t \bar t W^{+}$) and $4\%$ ($t \bar t W^{-}$) for $\mu_0 = Q$ and decreased by $0.9\%$ ($t \bar t W^{+}$) and $0.6\%$ ($t \bar t W^{-}$) for $\mu_0 = m_t + \frac{m_W}{2}$. The resummation reduces the scale uncertainty and brings the predictions for the two different scale choices closer together.

\begin{table}
 \centering
 \begin{tabular}{c|c|c|c|c}
  process & $\mu_0$ & NLO & NLO + NLL & NLO+NLL w $\mathbf{H}^{(1)}$\\\hline
  $t \bar t W^{+}$ & $Q$ & $329.9_{-11.1\%}^{+12.5\%}$\,fb & $332.1_{-11.2\%}^{+12.5\%}$\,fb & $341.1_{-8.6\%}^{+10.7\%}$\,fb \\
  $t \bar t W^{+}$ & $\frac{M}{2}$ & $422.1_{-11.5\%}^{+12.8\%}$\,fb & $423.5_{-11.5\%}^{+13.2\%}$\,fb & $418.4_{-10.0\%}^{+12.8\%}$\,fb\\
  $t \bar t W^{-}$ & $Q$ & $168.5_{-11.2\%}^{+12.7\%}$\,fb & $170.0_{-10.7\%}^{+12.1\%}$\,fb & $175.3_{-8.4\%}^{+9.9\%}$\,fb\\
  $t \bar t W^{-}$ & $\frac{M}{2}$ & $215.6_{-11.8\%}^{+13.4\%}$\,fb & $216.4_{-11.6\%}^{+13.8\%}$\,fb & $214.4_{-10.1\%}^{+13.4\%}$\,fb
 \end{tabular}
 \caption{Total $t \bar t W^{+/-}$ cross sections for the two different central scale choices and their scale uncertainty, which was calculated with the seven point method}
 \label{tab:xsections}
\end{table}

\acknowledgments
This work has been supported in part by the DFG grant KU 3103/1.
Support of the Polish National Science Centre grant no. DEC-2014/13/B/ST2/02486 is
gratefully acknowledged.
This work was also partially supported by the U.S. National
Science Foundation, under grants PHY-0969510, the LHC Theory Initiative, PHY-1417317
and PHY-1619867. 
TS acknowledges support in the form of the WWU Internationalisation scholarship.
DS thanks the organizers of the conference and the conveners of the top and electroweak session for the possibility to present this talk.


\newpage
\begin{thebibliography}{99}

\bibitem{Aaboud:2016xve}
  M.~Aaboud {\it et al.} [ATLAS Collaboration],
  \emph{Eur.\ Phys.\ J.\ C} {\bf 77} (2017) no.1,  40
  [{\tt arXiv:1609.01599 [hep-ex]}].

\bibitem{Khachatryan:2015sha}
  V.~Khachatryan {\it et al.} [CMS Collaboration],
  \emph{JHEP} {\bf 1601} (2016) 096
  [{\tt arXiv:1510.01131 [hep-ex]}].
  
\bibitem{Lazopoulos:2008de}
  A.~Lazopoulos, T.~McElmurry, K.~Melnikov and F.~Petriello,
  \emph{Phys.\ Lett.\ B} {\bf 666} (2008) 62
  [{\tt arXiv:0804.2220 [hep-ph]}].

\bibitem{Lazopoulos:2007bv}
  A.~Lazopoulos, K.~Melnikov and F.~J.~Petriello,
  \emph{Phys.\ Rev.\ D} {\bf 77} (2008) 034021
  [{\tt arXiv:0709.4044 [hep-ph]}].

\bibitem{Kardos:2011na}
  A.~Kardos, Z.~Trocsanyi and C.~Papadopoulos,
  \emph{Phys.\ Rev.\ D} {\bf 85} (2012) 054015
  [{\tt arXiv:1111.0610 [hep-ph]}].

\bibitem{Campbell:2012dh}
  J.~M.~Campbell and R.~K.~Ellis,
  \emph{JHEP} {\bf 1207} (2012) 052
  [{\tt arXiv:1204.5678 [hep-ph]}].

\bibitem{Alwall:2014hca}
  J.~Alwall {\it et al.},
  \emph{JHEP} {\bf 1407} (2014) 079
  [{\tt arXiv:1405.0301 [hep-ph]}].

\bibitem{Garzelli:2011is}
  M.~V.~Garzelli, A.~Kardos, C.~G.~Papadopoulos and Z.~Trocsanyi,
  \emph{Phys.\ Rev.\ D} {\bf 85} (2012) 074022
  [{\tt arXiv:1111.1444 [hep-ph]}].

\bibitem{Garzelli:2012bn}
  M.~V.~Garzelli, A.~Kardos, C.~G.~Papadopoulos and Z.~Trocsanyi,
  \emph{JHEP} {\bf 1211} (2012) 056
  [{\tt arXiv:1208.2665 [hep-ph]}].
  
\bibitem{Frixione:2015zaa}
  S.~Frixione, V.~Hirschi, D.~Pagani, H.-S.~Shao and M.~Zaro,
  \emph{JHEP} {\bf 1506} (2015) 184
  [{\tt arXiv:1504.03446 [hep-ph]}].

\bibitem{Li:2014ula}
  H.~T.~Li, C.~S.~Li and S.~A.~Li,
  \emph{Phys.\ Rev.\ D} {\bf 90} (2014) no.9,  094009
  [{\tt arXiv:1409.1460 [hep-ph]}].

\bibitem{Broggio:2016zgg}
  A.~Broggio, A.~Ferroglia, G.~Ossola and B.~D.~Pecjak,
  \emph{JHEP} {\bf 1609} (2016) 089
  [{\tt arXiv:1607.05303 [hep-ph]}].

\bibitem{Broggio:2017kzi}
  A.~Broggio, A.~Ferroglia, G.~Ossola, B.~D.~Pecjak and R.~D.~Sameshima,
  \emph{JHEP} {\bf 1704} (2017) 105
  [{\tt arXiv:1702.00800 [hep-ph]}].

\bibitem{Contopanagos:1996nh} 
  H. Contopanagos, E. Laenen and G. F. Sterman, 
  \emph{Nucl. Phys. B} {\bf 484} (1997) 303 
  [{\tt hep-ph/9604313}].

\bibitem{Kidonakis:1998nf}
  N. Kidonakis, G. Oderda and G. F. Sterman, 
  \emph{Nucl. Phys. B} {\bf 531} (1998) 365 
  [{\tt hep-ph/9803241}].

\bibitem{Catani:1996yz} 
  S. Catani, M. L. Mangano, P. Nason and L. Trentadue, 
  \emph{Nucl. Phys. B} {\bf 478} 273 (1996)
  [{\tt hep-ph/9604351}].

\bibitem{Bonciani:1998vc} 
  R. Bonciani, S. Catani, M. L. Mangano and P. Nason, 
  \emph{Nucl. Phys. B} {\bf 529} 424 (1998)
  [{\tt hep-ph/9801375}].

\bibitem{Kulesza:2015vda} 
  A. Kulesza, L. Motyka, T. Stebel and V. Theeuwes, 
  \emph{JHEP} 1603 (2016) 065 
  [{\tt arXiv:1509.02780 [hep-ph]}].

\bibitem{Beenakker:2011sf} 
  W. Beenakker, S. Brensing, M. Kramer, A. Kulesza, E. Laenen and I. Niessen, 
  \emph{JHEP} 1201 (2012) 076
  [{\tt arXiv:1110.2446 [hep-ph]}].

\bibitem{Beenakker:2013mva} 
  W. Beenakker et al., 
  \emph{JHEP} 1310 (2013) 120 
  [{\tt arXiv:1304.6354 [hep-ph]}].

\bibitem{pdf} 
  L. A. Harland-Lang, A. D. Martin, P. Motylinski and R. S. Thorne, 
  \emph{Eur. Phys. J. C} {\bf 75} (2015) 5, 204
  [{\tt arXiv:1412.3989 [hep-ph]}].

\bibitem{Kulesza:2017ukk}
  A.~Kulesza, L.~Motyka, T.~Stebel and V.~Theeuwes,
  [{\tt arXiv:1704.03363 [hep-ph]}].
\end{thebibliography}
\end{document}